# Structural, elastic, electronic, magnetic and thermoelectric properties of new quaternary Heusler compounds CoZrMn*X* (*X*=Al, Ga, Ge, In)

M. Anwar Hossain*, Md. Taslimur Rahman, Morioum Khatun and Enamul Haque

Department of Physics, Mawlana Bhashani Science and Technology University,

Santosh,Tangail-1902, Bangladesh

Email: anwar647@mbstu.ac.bd

## Abstract

We have performed a comprehensive set of first principles calculations to study the structural, elastic, electronic, magnetic and transport properties of new quaternary Heusler compounds CoZrMn*X* (*X* =Al, Ga, Ge, In). The results showed that all the quaternary Heusler compounds were stable in Type(I) structure. CoZrMn*X* are elastically stable and relatively hard materials. CoZrMnAl, CoZrMnGa, and CoZrMnIn are found to be ductile and CoZrMnGe is brittle in nature. The calculated Debye temperatures of all compounds are relatively high. The electronic structure calculations reveal that CoZrMnAl is nearly half metallic, CoZrMnGa and CoZrMnIn are metallic, and CoZrMnGe is a narrow indirect bandgap semiconductor. The calculated magnetic properties implies that CoZrMnAl, CoZrMnGa, and CoZrMnIn are ferromagnetic while CoZrMnGe is non-magnetic material. The CoZrMnAl is highly spin-polarized (96%) and CoZrMnGe is non-spin-polarized. Seebeck coefficent (*S*) in CoZrMnGe is relatively high (-106 µV/K at 650K) due to its semiconducting nature. The calculated thermoelectric figure of merit CoZrMnGe is 0.1 at 600K and for CoZrMnIn it is also 0.1 at 900 K. We hope our interesting results will inspire experimentalist to synthesis the new quaternary Heusler compounds CoZrMn*X* (*X* =Al, Ga, Ge, In).

**Keywords:** Heusler compound; Electronic structure; Elastic properties; Magnetic properties; Thermoelectric properties

## 1.      Introduction

The materials with high spin-polarization, high Curie temperature, and half-metallicity have great importance for different technological applications and Heusler compounds can exhibit these important properties [1–7]. Thus, these materials have brought a great research interest to discover new compounds exhibiting the above properties [8–12]. Heusler compounds with these properties may be used in the wide range of applications such as spin injectors [13], spin valves [14], magnetic tunnel junctions [15], and spin torque based transfer-random access memories [16]. In half-metallic Heusler compounds, the top edge and bottom edge of the highest and the lowest, respectively, occupied spin-down bands, touch the Fermi level, and destroy the indirect gap [3]. These type of nature of bands depend on the total valence electrons per unit cell of the compound. The Heusler compounds also exhibit good thermoelectric properties [17–19] and superconductivity [20]. The $Hf_{0.6}Zr_{0.40}NiSn_{0.98}Sb_{0.02}$ Heusler alloy exhibits a high thermoelectric figure of merit (ZT) 1.0 at 1073 K [17]. The most of the explored magnetic Heusler compounds are ternary intermetallics $X_2Y'Z$ ( where $X$ and $Y'$ are transition metals and $Z$ is the main group element) and the presence of unoccupied d states of magnetic elements in sub-lattices belonging to different symmetries gives rise to the magnetic properties.

The ternary intermetallics, $X_2Y'Z$, usually has the $L2_1$ structure with the space group of $Fm\overline{3}m$ (#225). When one of the X atoms is replaced by another transition metal element, we can find the structure of a quaternary Heusler compound $XX'YZ$ with the space group of $F$-43$m$ (space group 216). It is found that sub-lattice are possible in $XX'Y'Z$ compounds, where $X'$ is a magnetic element different from $X$, more than that are possible with $X_2Y'Z$ materials due to the occupancy of each sub-lattice with different elements [21]. Recently, many studies have been reported on the quaternary Heusler compounds ($XX'Y'Z$) exhibiting half-metallicity [22,23], high Curie temperature [21,24] and spin gapless semiconducting properties [25] and found to be suitable for spintronic device applications. The quaternary Heusler compounds ($XX'Y'Z$) with a combination of 3d and 4d elements in the periodic table exhibit high Curie temperature [26,27]. These compounds can be formed in three types of structure. In all types, the $Z$ atoms are fixed at the 4a (0, 0, 0) Wyckoff positions. In type-I (T-I) structure, $X'$ atoms occupy at the 4b (0.5, 0.5, 0.5) Wyckoff positions, $X$ at the 4c (0.25, 0.25, 0.25) and $Y'$ atoms at the 4d (0.75, 0.75, 0.75) Wyckoff positions. In type-II (T-II) structure, $X'$ and $Y'$ have interchanged their positions and in type-III, $X'$ and $X$ are interchanged.

In this paper, the structural, elastic, electronic, magnetic and transport properties of new quaternary Heusler compounds CoZrMn$X$ ($X =$Al, Ga, Ge, In) have been studied by using the first-principles calculations. We have chosen this series of compounds with the hope that they would exhibit half-metallicity, good magnetic and thermoelectric properties. Our studies indicate that all compounds are stable and CoZrMnAl has been found to be half-metallic ferromagnetic while CoZrMnGe has been found to be narrow bandgap semiconductor. Although CoZrMnIn has been found to be metallic, it exhibits high thermoelectric figure of merit ZT (0.1) among the four compounds.

## 2.      Methods of calculations

The structural, electronic, and magnetic properties were studied by using full potential linearized augmented plane wave (LAPW) method as implemented in WIEN2k [28]. The muffin tin radii: 2.46, 2.34, 2.46 Bohr for Co, Zr, Mn, respectively and 2.22, 2.25, 2. 38, 2.44, 2.5 Bohr for Al, Ga, In, Ge, respectively, have been used. We have calculated the formation energies of these compounds by using plane wave pseudopotential method as implemented in Quantum espresso [29]. The calculations have been performed by using 408 eV kinetic energy cutoff for wavefunctions and 3265 eV for charge density. The elastic constants and related parameters were calculated by using IRelast method [30]. The magnetic properties were studied by performing spin-polarization calculations without including spin-orbit coupling effect. The PBE-GGA [31,32] functional has been used. Thus, we have studied electronic and magnetic properties of the four compounds in WIEN2k. The convergence criteria have been set to $10^{-4}$Ry for energy and 0.001e for charges in WIEN2k. The dense mesh of $21 \times 21 \times 21$ k-point has been used to perform the above calculations. The thermoelectric transport properties were calculated by solving semi-classical Boltzmann transport equation as implemented in BoltzTraP [33]. A denser mesh of $43 \times 43 \times 43$ k-points has been used in WIEN2k to study transport properties and generate the required input files for BoltzTraP. The chemical potential has been considered to be equal to the zero temperature Fermi energy for transport properties calculation.

## 3.      Result and Discussions

### 3.1.    Structural properties

The variation of energy with the volume of the unit cell in per formula unit are illustrated in Fig. 1 for all possible structure types. From the Fig. 1, it is clear that all the four compounds have the smallest energy in the type-I structure than other two type-structure. Thus, type-I (T-I) structure is the most stable structure for our considered compounds. The stability of these compounds can be predicted by calculating the formation energy. The formation energy of a material can be calculated by the following equation:

$$E_F = E_{CoZrMnX} - (E_{Co} + E_{Zr} + E_{Mn} + E_X) \tag{1}$$

where $E_{CoZrMnX}$ is the total ground state energy of CoZrMn$X$ compounds per formula unit, $E_{Co}$, $E_{Zr}$, $E_{Mn}$, and $E_X$ are the total ground state energies of the bulk Co, Zr, Mn, and $X$ ($X$=Al, Ga, Ge, In) crystals. The calculated formation energies and optimized lattice parameters are listed in the Table-1. The negative enthalpy formation energy implies that the material is thermally stable. The optimized lattice parameters are comparable with the experimental and theoretical values of other compounds with the same structure [21,23,24].

Table-1: The optimized lattice parameters and formation energy of CoZrMn$X$.

| Compounds | Lattice parameters a (Å) | Formation energy f (eV) | Type of structure |
|-----------|--------------------------|-------------------------|-------------------|
| CoZrMnAl | 6.0746 | -3.1349 | T-I |
| CoZrMnGa | 6.0685 | -1.2481 | T-I |
| CoZrMnGe | 6.0512 | -2.4348 | T-I |
| CoZrMnIn | 6.2752 | -1.6794 | T-I |

The negative value of formation energy of all compounds indicates that they are thermally stable and their chemical synthesis would be possible. However, among the quaternary Heusler compounds, the only CoRuFeSi has been synthesized [24]. The experimental studies are needed to synthesize these compounds and reveal the possibility of formation of other theoretically predicted compounds. The elastic properties reflect the mechanical stability of a material and hence is important for different practical usages. The calculations of elastic constants and moduli of elasticity of a cubic crystal are straight forward and can be found in the standard article[34–36]. Our calculated elastic constants, elastic moduli, and Poisson's ratio of CoZrMn*X* are presented in Table-2.

Table-2: Calculated elastic constants, elastic moduli in (GPa), and Poisson's ratio.

| Compounds | $C_{11}$ | $C_{12}$ | $C_{44}$ | $B$ | $G$ | B/G | $E$ | Y |
|---|---|---|---|---|---|---|---|---|
| CoZrMnAl | 261.1437 | 115.4068 | 85.8098 | 163.985 | 80.375 | 2.04 | 207.262 | 0.289 |
| CoZrMnGa | 214.9019 | 108.7590 | 82.9659 | 144.139 | 69.358 | 2.07 | 179.312 | 0.292 |
| CoZrMnGe | 238.3834 | 65.3708 | 91.3630 | 123.041 | 89.387 | 1.38 | 215.882 | 0.207 |
| CoZrMnIn | 209.6486 | 117.0365 | 86.1891 | 147.906 | 67.169 | 2.20 | 175.013 | 0.302 |

The necessary and sufficient mechanical stability conditions of the elastic constants for a cubic crystal are[37]

$$C_{11} - C_{12} > 0, \ C_{44} > 0, (C_{11} + 2C_{12}) > 0 \tag{2}$$

Our calculated elastic constants satisfy the above conditions and therefore, all of our theoretically predicted compounds CoZrMn*X*are elastically stable.

The Pugh's (B/G) ratio [38], Cauchy pressure ($c_{11}$-$c_{44}$) [39] and Poisson's ratio limit suggest that CoZrMnGe compound is brittle (but other three CoZrMnZ (Z=Al, Ga, In) are ductile).The shear anisotropy factor can be calculated from the equation [40], $A = 2C_{44}/(C_{11} - C_{12})$. The hardness of a material is important for practical applications and can be predicted by using the equation [41], $H_V = 2\left(\left(\frac{G}{B}\right)^2 G\right)^{0.585} - 3$. The calculated shear anisotropic factor and Vickers hardness of

CoZrMn$X$ are given in the Table-3.

Table 3: Calculated elastic shear anisotropic factor $A$, and hardness (in GPa).

| Compounds | $A$ | $H_V$ |
|-----------|-------|--------|
| CoZrMnAl | 1.178 | 8.303 |
| CoZrMnGa | 1.563 | 7.148 |
| CoZrMnGe | 1.056 | 16.062 |
| CoZrMnIn | 1.861 | 6.307 |

The values of Vickers hardness indicate that CoZrMn$X$ are relatively hard material and CoZrMnGe is harder than others compounds. All compounds show high shear anisotropic nature. The Debye temperature, $\Theta_D$, related to thermal stability of a material, can be calculated by using the following equation [42]

$$\Theta_D = \frac{h}{K_B}\left(\frac{3N\rho}{4\pi M}\right)^{1/3} v_m \qquad (3)$$

where $\rho$, $v_m$ are density, and average sound velocity, respectively. The average sound velocity of material can be expressed as, $v_m = \left[\frac{1}{3}\left(\frac{2}{v_t{}^3} + \frac{1}{v_l{}^3}\right)\right]^{-1/3}$, where $v_t$ and $v_l$ stand for the transverse and longitudinal sound velocity, respectively, and can be obtained from $v_l = \left(\frac{3B+4G}{3\rho}\right)^{1/2}$ and $v_t = \left(\frac{G}{\rho}\right)^{1/2}$. The melting temperature is important factor for material used in the heating system and can be theoretically predicted by the equation [10] $T_m = \left[553\text{K} + \left(\frac{5.91\text{K}}{GPa}\right)C_{11}\right] \pm 300$ K. Our calculated Debye temperature and predicted melting temperature of CoZrMn$X$ are listed in the Table-4. It is clear from the table that the Debye temperatures of all compounds are in the range of values that than obtained for CoRuFeZ ($Z$=Si, Ge, Sn) Heusler compounds [8].

Table-4: Vibrational properties and predicted melting temperature $T_m$ (K) from elastic constants: $v_l$, $v_t$, and $v_m$ (in m/s), $\Theta_D$ (in K).

| Compounds | $v_t$ | $v_l$ | $v_m$ | $\Theta_D$ | $T_m \pm 300$ |
|---|---|---|---|---|---|
| CoZrMnAl | 3435.61 | 6310.3 | 3832.37 | 471.75 | 2096.40 |
| CoZrMnGa | 2911.98 | 5378.52 | 3249.61 | 401.92 | 1823.10 |
| CoZrMnGe | 2292.71 | 5420.32 | 3638.07 | 449.57 | 1961.85 |
| CoZrMnIn | 2820.77 | 5303.74 | 3151.85 | 374.50 | 1792.02 |

## 3.2. Electronic properties

The calculated energy band structures of CoZrMn$X$ are presented in Fig. 2 without considering spin effect. The most of the conduction bands of CoZrMn$X$ are non-dispersive. These conduction bands are also non-flat. The plotted energy band structures of CoZrMnAl, CoZrMnGa, and CoZrMnIn are mainly d-orbital character. These energy bands arise from the Co-3d, and Mn-3d states with a small contribution of Zr-4d. Due to the d-like character of energy band structure, the polarization in these compounds should be high [43]. When the spin effect is considered in the band structure calculation, CoZrMnAl band structure for spin-down shows overlapping between conduction and valence bands at the Fermi while for spin-up, two valence bands at X-point cross the Fermi level as shown in the Fig. S2(a) (in supplementary). Therefore, CoZrMnAl can be regarded as nearly half-metallic (see also the density of states, Fig. S2(b)). The band structure of CoZrMnGa and CoZrMnIn for the majority and minority carriers shows almost identical feature (see Figs. S4 and S8). The overlapping is small for majority carriers than that for minority carriers. But these compounds exhibit metallic nature for both spin. The band structure of CoZrMnGe is a mixed d+p-like character. The conduction and valence bands of CoZrMnGe do not overlap at the Fermi level (see Fig. 2 (c)) and remain 0.042 eV bandgap between them at Γ and X-points. Therefore, CoZrMnGe is a narrow indirect bandgap semiconductor. The inclusion of spin effect on the band structure calculation of CoZrMnGe has no effect and band structure for the majority and minority carriers is identical (see Fig. S6 in the supplementary). The total and projected density of states (DOS) of CoZrMn$X$ are illustrated in the Figs. 3-6 without including spin effect, for spin effect, see Figs. S2-S9 in the supplementary). The main feature of the density of states of CoZrMnAl, CoZrMnGa, and CoZrMnIn are almost identical. The Co-3d, and Mn-3d states have the dominant contributions to the total density of states at the Fermi level. The Zr-4d states have a small contribution to the total density of states. The highest peak at the energy ~-1.8eV arises from the strong hybridization between Co-3d and Al-3p orbitals (see Fig. 3). The corresponding peak at the energy ~1.65 eV represent the corresponding antibonding combinations. Similarly, the peak at the energy ~-1.56 eV comes from the sigma bonding combinations between Co-3d and Ga-3p states as shown in the Fig. 4 (a). The feature of the density of states of CoZrMnGe is totally different from other compounds as shown in the Fig. 5. The density of states at the Fermi level of CoZrMnGe is zero and the semiconducting nature arises from the strong hybridization  Co-3d and Ge-4p states. The highest peak at the energy ~-0.8 eV comes from the sigma bonding combinations of Mn-3d and Ge-4p states. The peak at positive energy corresponds to the antibonding combinations. The density of states calculated by considering spin effect is same for both majority and minority carriers (see Fig. S5).Thus, the spin-polarization is zero for CoZrMnGe, as given in the Table-6. The Co-3d, Mn-3d orbitals have the dominant contributions to the total density of states of at the Fermi level of CoZrMnIn as illustrated in the Fig. 6. The bonding and antibonding combinations of Co-3d and In-3p states are similar to that CoZrMnAl, and CoZrMnGa. Below the Fermi level, a large pseudogap exists for all compounds. See supplementary document for further details and spin-polarization calculations of band structure and DOS.

### 3.3. Magnetic properties

The magnetic moment determines the magnetic nature of a material. The calculated atomic and total magnetic moment of CoZrMn$X$ are listed in the Table-5. The magnetic moment of CoZrMnAl ($1.0030\mu_B$>1) indicates that CoZrMnAl is ferromagnetic at 0 K. The CoZrMnGa and CoZrMnIn compounds are also ferromagnetic and their magnetic moments are larger than that for CoZrMnAl. The magnetic moment of CoZrMnGe is very small and thus CoZrMnGe is non-magnetic material. It is note that Co and Mn atoms have the prominent contributions to the total magnetic moment of CoZrMn$X$. The Zr and $X$ atoms do not favor for magnetism in all compounds. The calculated magnetic moments of CoZrMn$X$ do not properly fit with Slater-Pauling rule[44,45] of half-metallicity, $M = N_V - 24$, where $N_v$ is the number of valence electrons in a unit cell and M is the magnetic moment. It is expected, any compound of CoZrMn$X$ is not completely half-metallic. The variations of total magnetic moment with lattice parameters is illustrated in the Fig. 7. In all cases, the magnetic moment increases with lattice parameters. As magnetic moment goes to high, the variations gradually become smother, as expected[46]. At small lattice parameters (<6.0Å), CoZrMnAl becomes paramagnetic.

Table-5: Calculated atomic and total magnetic moment of CoZrMn$X$ at equilibrium lattice parameters.

| Compounds | Atomic magnetic moment in the unit cell | | | | Total magnetic |
|---|---|---|---|---|---|
| | Co | Zr | Mn | $X$ | moment ($\mu_B$) |
| CoZrMnAl | 0.046 | -0.0602 | 1.04 | -0.0270 | 1.0030 |
| CoZrMnGa | 0.7509 | -0.1888 | 2.4658 | -0.0348 | 2.9182 |
| CoZrMnGe | 0.0027 | -0.0006 | 0.0053 | -0.00003 | 0.0075 |
| CoZrMnIn | 1.042 | -0.2235 | 3.0384 | -0.02126 | 3.7579 |

These properties implies that the magnetic moment is strongly dependent on the structure of the material, i.e., lattice parameters. The spin-polarization property at the Fermi of a material can be calculated by the using the density of states at the Fermi level for the corresponding spin defined as [43]

$$P = \frac{N_\uparrow(E_F) - N_\downarrow(E_F)}{N_\uparrow(E_F) + N_\downarrow(E_F)} \qquad (4)$$

where $N_\uparrow(E_F)$ and $N_\downarrow(E_F)$ are the density of states at the Fermi level for spin-up and spin-down, respectively. If the spin-up(or spin-down) contribution to the density states at the Fermi level is zero, a material will be 100% spin-polarized. For the d-like orbital character, P will be high but low for s-like or s+d-like orbital character [43]. The calculated spin polarized total and projected density of states of CoZrMn$X$ are shown in the Figs. S2-S9 (see supplementary document for these figures). The calculated spin-polarizations (P) of CoZrMn$X$ are listed in the Table-6.

Table-6: The calculated density of sates at the Fermi level for spin-up $N_\uparrow(E_F)$ (states/ eV f.u.) and spin-down $N_\uparrow(E_F)$ (states/ eV f.u.), as well as the polarization in %.

| Compounds | $N_\uparrow(E_F)$ | $N_\downarrow(E_F)$ | P (%) |
|---|---|---|---|
| CoZrMnAl | 0.085 | 4.16 | 96 |
| CoZrMnGa | 1.76 | 6.77 | 59 |
| CoZrMnGe | 0 | 0 | 0 |
| CoZrMnIn | 4.31 | 1.39 | 51 |

The density of states for majority carriers of CoZrMnAl is very small and for minority carriers is large, making the material to be highly spin-polarized. Therefore, CoZrMnAl is nearly half metallic ferromagnetic material. The CoZrMnGa and CoZrMnIn compounds exhibit relatively smaller spin-polarization due to the contributions of both majority and minority carriers. CoZrMnGe is non-spin-polarized as expected since it is non-magnetic material. The highly spin-polarized materials are very important for spintronic device applications.

### 3.4. Transport properties

The thermoelectric transport properties of CoZrMn$X$ compounds are shown in the Fig. 8. Although CoZrMnAl is found to be half-metallic, Seebeck coefficient is small (33.05 μV/K at 1100 K) in this compound due to the non-flat and non-dispersive band. Another fact that the density of states at the Fermi level is high. The Seebeck coefficient (S) of CoZrMnGe decrease sharply up to 600 K and after then increases (see Fig. 8(a)). Below room temperature, the Seebeck coefficient is positive and from 300 K, it becomes negative. This indicates that CoZrMnGe remains p-type material below 300 K and becomes n-type above 300 K. At the temperature of 50 K, the S is very high (300 µV/K) and at 650 K, the maximum S is -106 µV/K. Semiconducting nature with small bandgap gives rise to such high thermopower in CoZrMnGe. The thermopower (S) of CoZrMnAl, CoZrMnGa, and CoZrMnIn is comparatively small and slowly increases with temperature up to 800 K and then slowly decreases. The CoZrMnGa, CoZrMnGe, and CoZrMnIn are p-type materials (as S is positive). The electrical conductivity (σ/τ) of CoZrMnAl decreases with temperature as shown in the Fig. 8(b). This implies the metallic nature of CoZrMnAl. The electrical conductivity of CoZrMnGa also exhibits similar trend implying metallic nature. However, this decreases with the temperature of CoZrMnGe as expected (since CoZrMnGe is semiconductor). The electrical conductivity of CoZrMnIn also implies its metallic nature. The electronic part of the thermal conductivity increases with temperature for all compounds as illustrated in the Fig. 8(c). The electronic part of the thermal conductivity of CoZrMnAl is highest and for CoZrMnGe, it is lowest among these compounds. The high thermal conductivity in CoZrMnAl arises from the combined contribution of electrons and phonons. The thermoelectric performance is characterized by dimensionless figure of merit Z defined as [47] $ZT = \frac{S^2\sigma}{\kappa}T$. The calculated thermoelectric figure of merit of CoZrMn$X$ at different temperature is presented in the Fig. 8(d). The maximum thermoelectric figure of merit for CoZrMnGe and CoZrMnIn is around 0.1.

## 4. Conclusions

In summary, we have predicted a series of new quaternary Heusler compounds CoZrMn$X$ ($X$ =Al, Ga, Ge, In) and studied the structural, elastic, electronic, magnetic and transport properties by using first-principles calculations. We have found that these compounds are thermally stable from point of view of enthalpy formation. These predicted Heusler compounds (CoZrMn$X$) are also elastically stable, and relatively hard materials. The CoZrMnAl, CoZrMnGa, and CoZrMnIn are found to be ductile and CoZrMnGe is brittle in nature. The calculated Debye temperatures of these compounds are relatively high. The electronic structure calculations reveal that CoZrMnAl is nearly half metallic, CoZrMnGa and CoZrMnIn are metallic, and CoZrMnGe is a narrow indirect bandgap (0.042 eV) semiconductor. The band structure of CoZrMnAl, CoZrMnGa, and CoZrMnIn are mainly d-like character while d+p-like character for CoZrMnGe. The calculated magnetic properties implies that CoZrMnAl, CoZrMnGa, and CoZrMnIn are ferromagnetic while CoZrMnGe is non-magnetic material. The CoZrMnAl is highly spin-polarized (96%). The Seebeck coefficient ($S$) in CoZrMnGe is relatively high (-106 μV/K at 650K) due to its semiconducting nature. The maximum thermoelectric figure of merit for CoZrMnGe is 0.1 at 600 K and for CoZrMnIn is 0.1 at 900 K. Thus, these two materials are potential candidates for thermoelectric applications. The thermoelectric performance of these materials could be improved by doping with suitable elements.

Figures

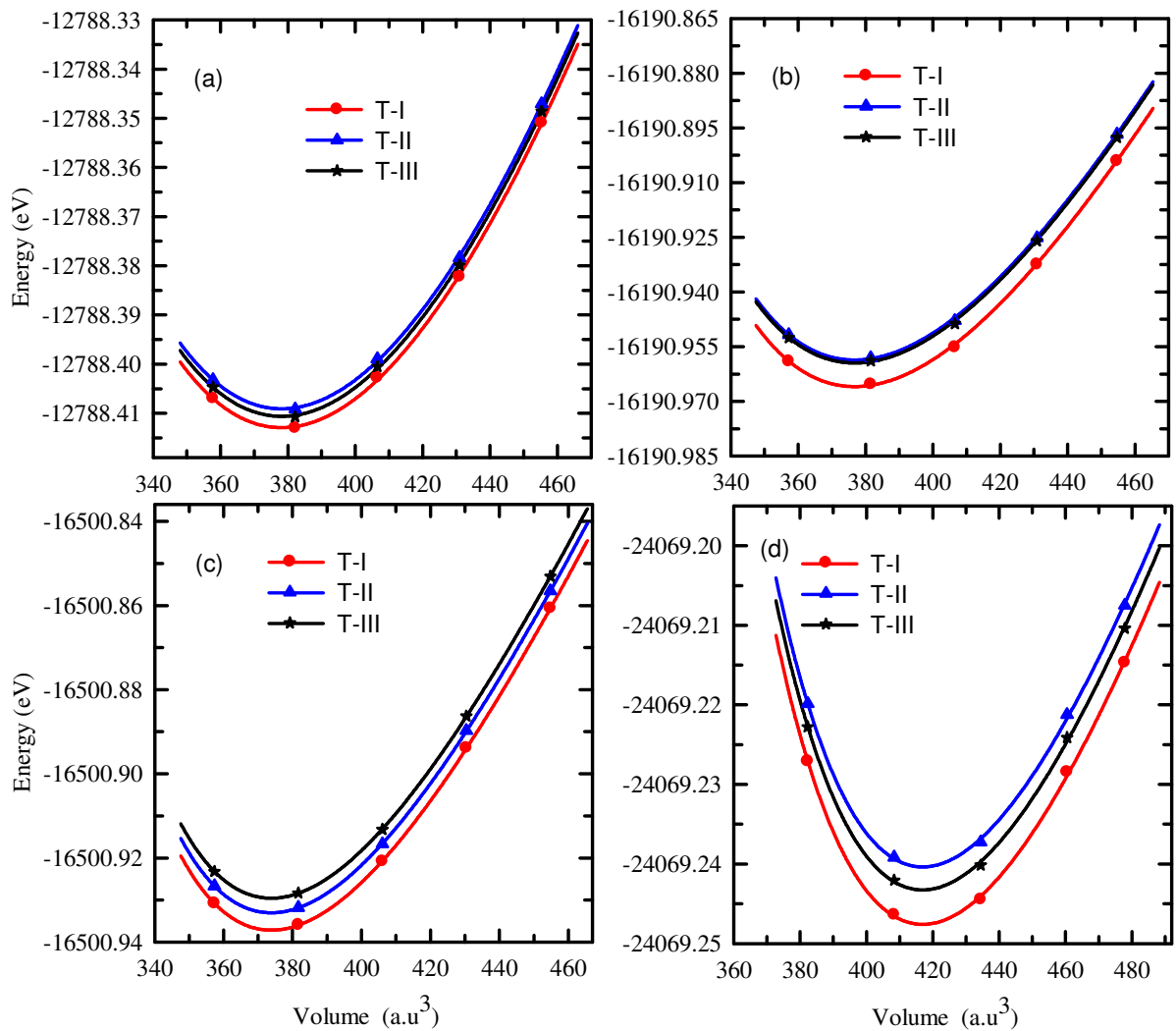

Fig. 1: Variations of energy with thevolume of four compounds in the tree type structures; the symbol '*T*' represents 'type'. The solid lines represent the data obtained by solving the Murnaghan equation of state and symbols represent SCF (self-consistent field calculation) data.

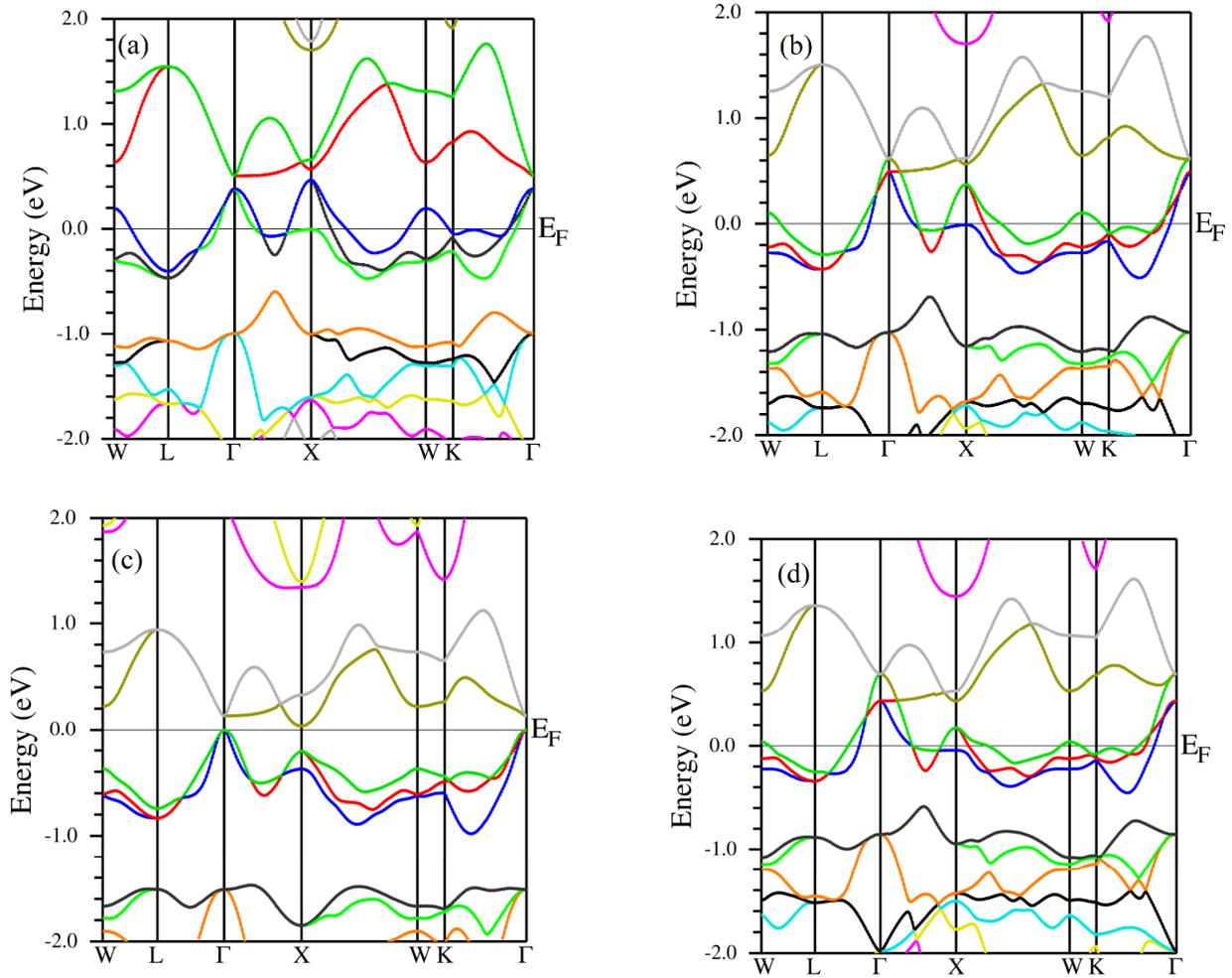

Fig. 2: The energy band structures of CoZrMn*X* by using PBE functional: (a) CoZrMnAl, (b) CoZrMnGa, (c) CoZrMnGe, and (d) CoZrMnIn. The horizontal line at the zero energy represents the Fermi level.

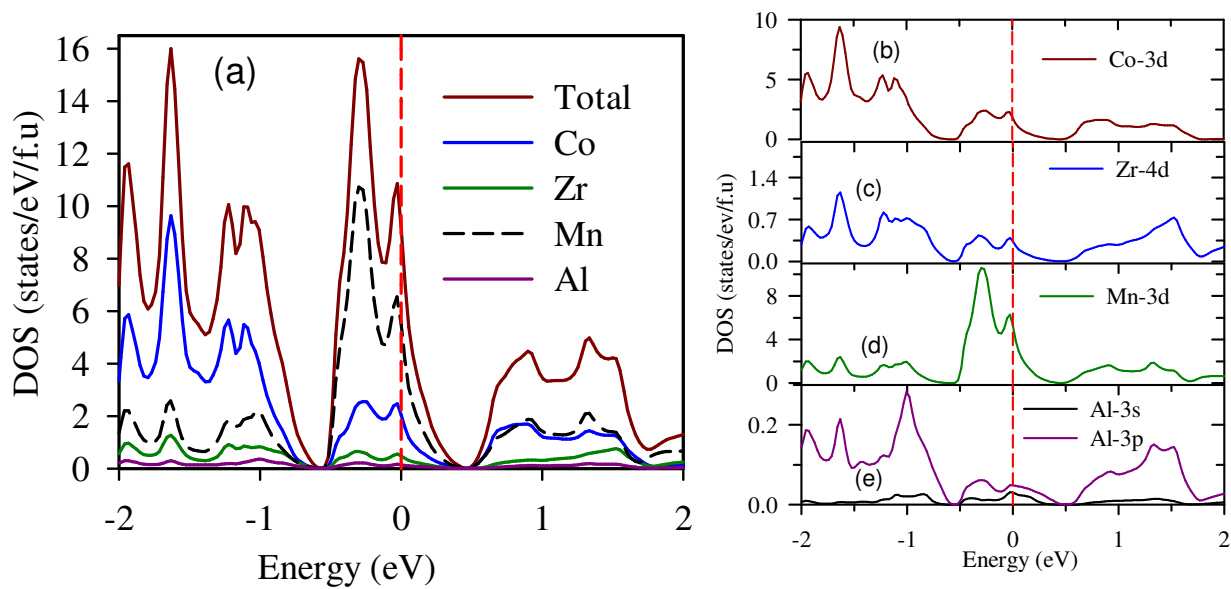

Fig. 3: Total and projected density of states of CoZrMnAl. The vertical dash line (red) at the zero energy represents the Fermi level.

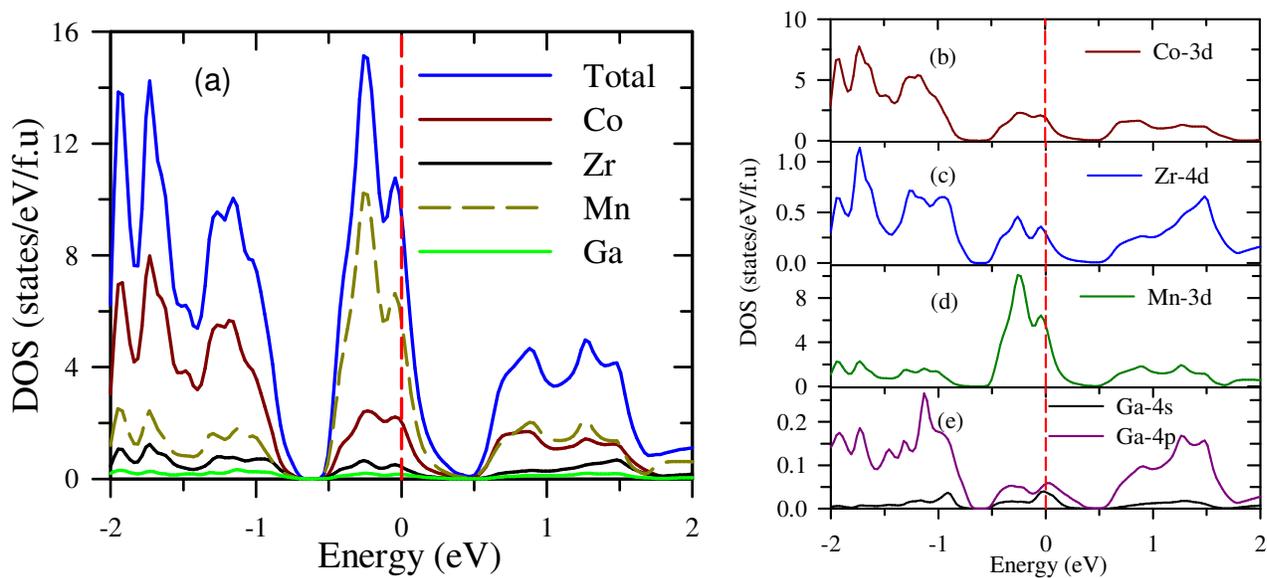

Fig. 4: Total and projected density of states of CoZrMnGa. The vertical dash line (red) at the zero energy represents the Fermi level.

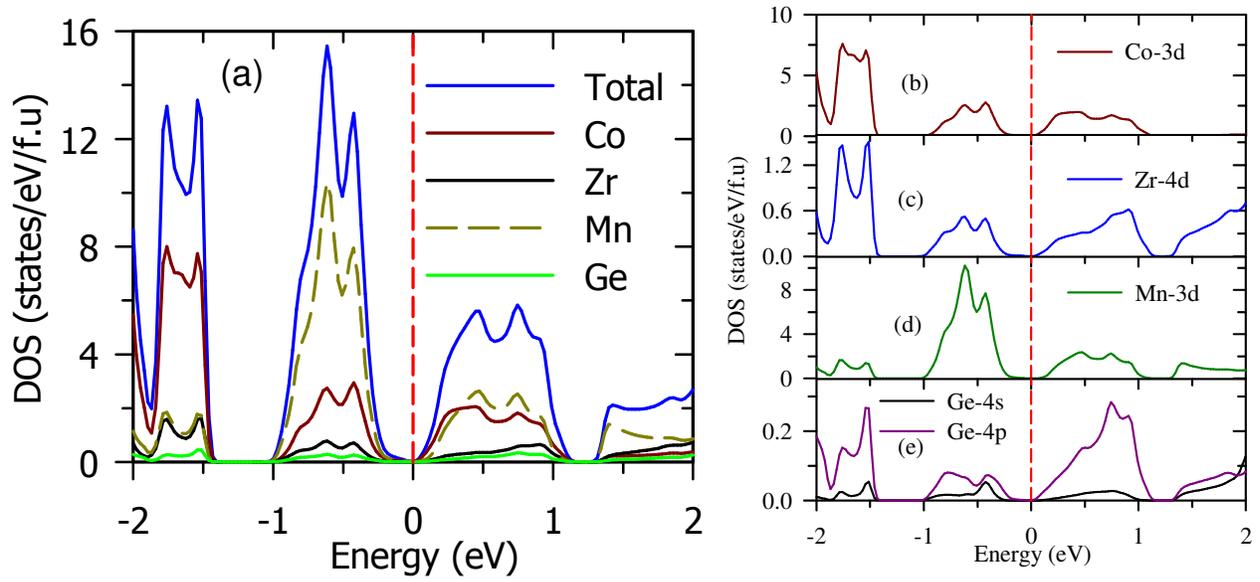

Fig. 5: The calculated total and projected density of states of CoZrMnGe. The vertical dash line (red) at the zero energy represents the Fermi level.

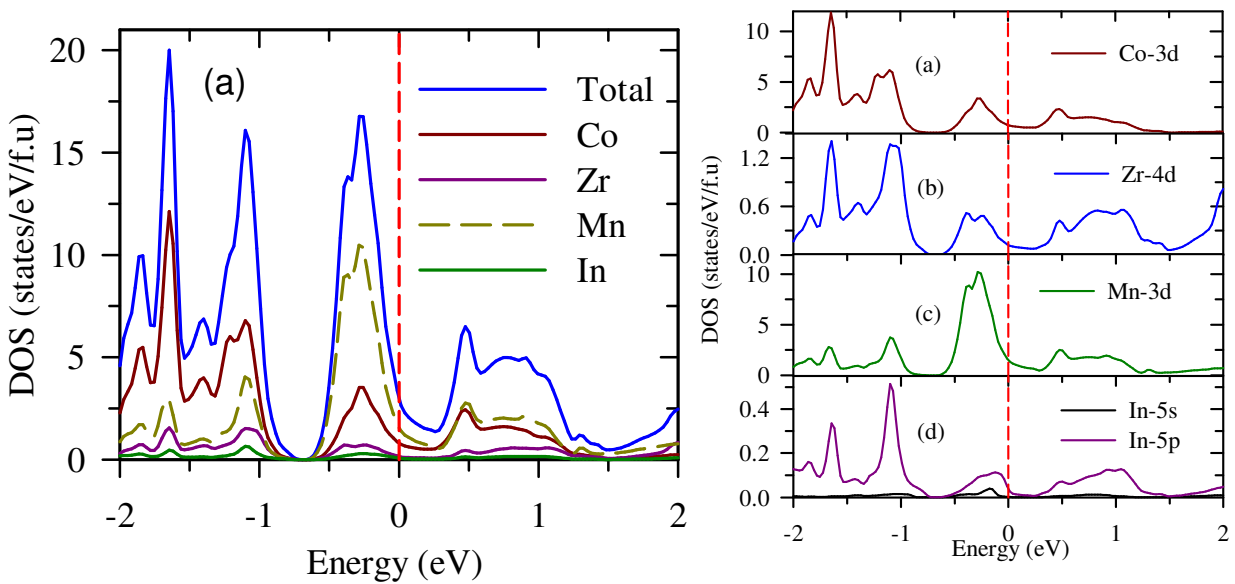

Fig. 6: Total and projected density of states of CoZrMnIn. The vertical dash line (red) at the zero energy represents the Fermi level.

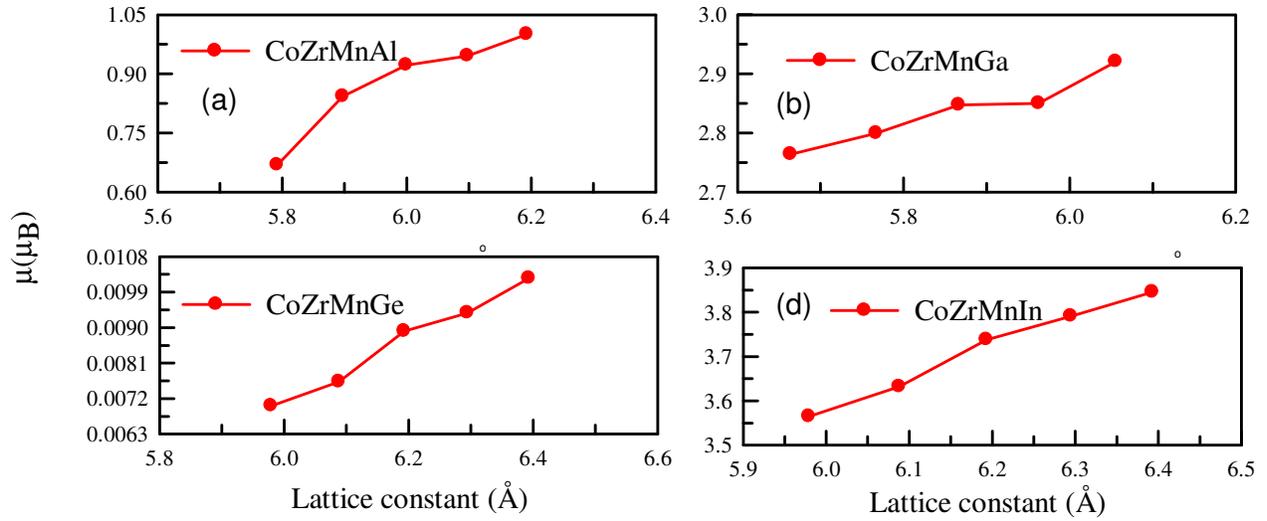

Fig. 7: The variations of thetotal magnetic moment in the cell per formula unit with the lattice parameters ofCoZrMn*X*.

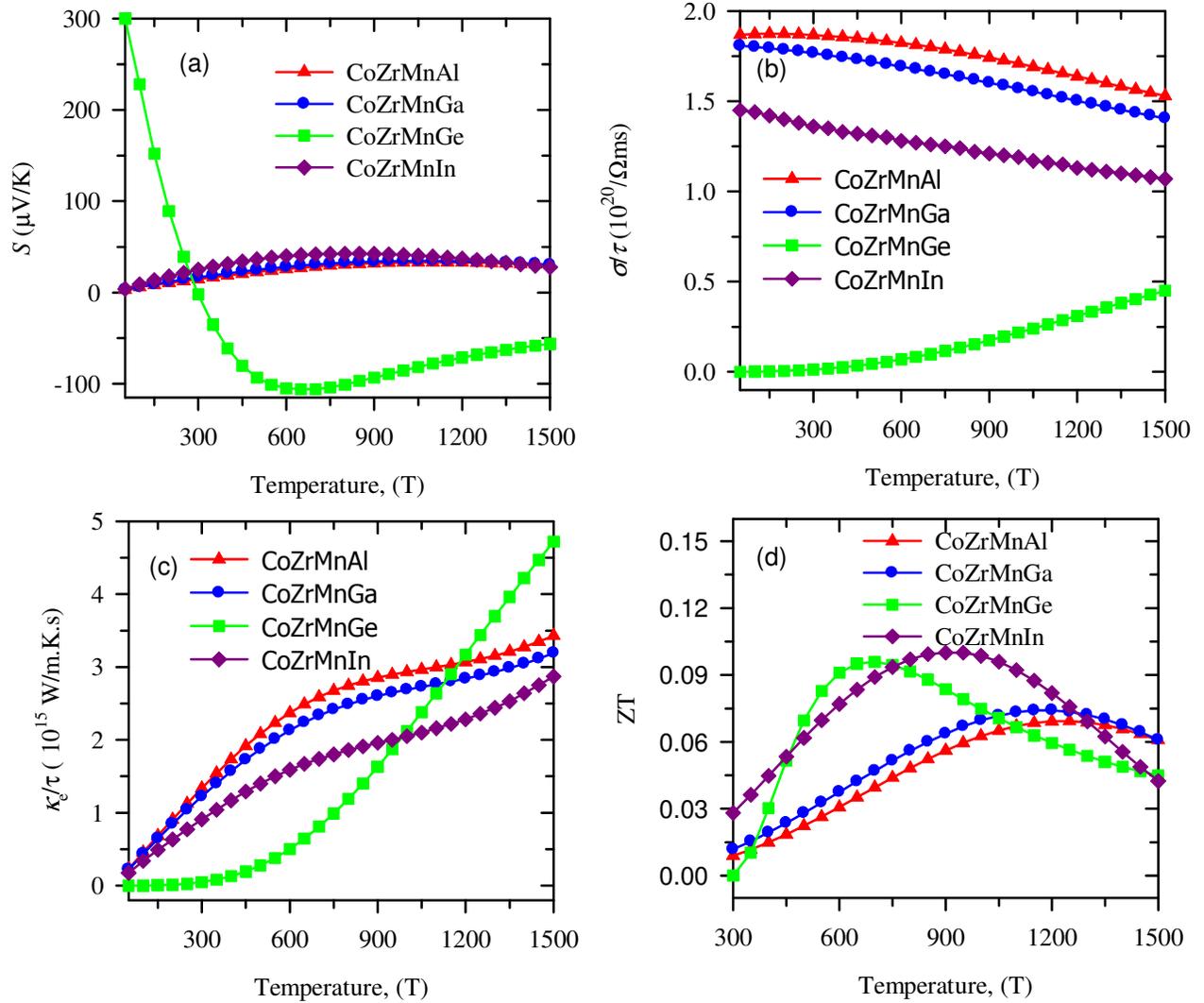

Fig. 8: Temperature dependence of thermoelectric transport properties of CoZrMn*X*. (a) Seebeck coefficient (*S*), (b) electrical conductivity (σ/τ), (c) electronic part of the thermal conductivity (κ$_e$/τ), (d) thermoelectric figure of merit (ZT).

# Supplementary Document for

## Structural, elastic, electronic, magnetic, and thermoelectric properties of new quaternary Heusler compounds CoZrMn*X* (*X*=Al, Ga, Ge, In).

**Introduction**

In this Supplementary, we have presented the details of thecomputational setup and figures of energy band structure and density of states for spin effect. The new compounds CoZrMn*X* (*X*=Al, Ga, Ge, In) have been found to exhibit some interesting properties such as nearly half-metallicity, semiconductivity, ferromagnetism, etc.. These are relatively hard materials and the Debye temperatures of them are high. Recently, many studies have been reported on the quaternary Heusler compounds[1,2].

**Computational setup**

Firstly, we have chosen a series of elements (*X*=Al, Ga, In Ge) from theperiodic table. Using the experimental lattice parameters of CoRuFeSi (5.773 Å) [3] as the reference values, the geometry optimization of these compounds have been performed in WIEN2k [4]. After optimization, the elastic properties of these compounds have been calculated by using IRelast script [5] in WIEN2k. The four compounds CoZrMn*X* (*X*=Al, Ga, Ge, In) have been found to be elastically stable.

Then we have studied the formation energies of these seven compounds by using plane wave pseudo potential method as implemented in Quantum espresso[6]. Finally, we have studied electronic and magnetic properties of four compounds in WIEN2k. The spin-polarization calculations have been performed to study magnetic properties. The face-centered cubic crystal structure of CoZrMn*X* is shown in the Fig. S1.

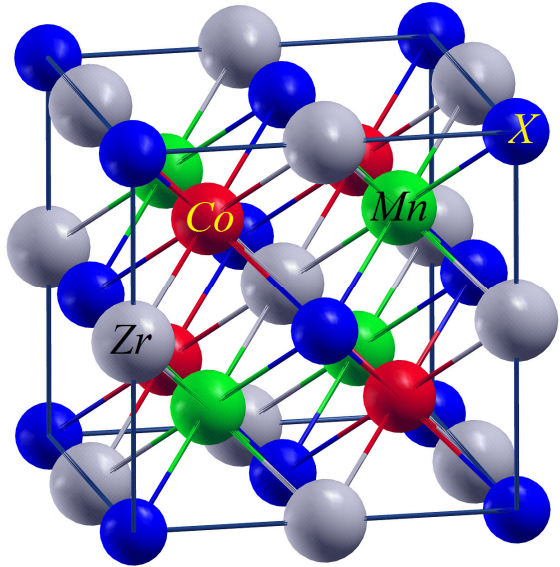

Fig. S1: Crystal structure of CoZrMn*X* (*X*=Al, Ga, Ge, In). This figure is plotted by using Xcrysden [9].

The crystal structure of CoZrMn*X* compounds contain large interatomic bonds. In CoZrMn*X*compounds, total 14 *X* (*X*=Al, Ga, Ge, In) atoms exist.

**Electronic properties**

The electronic properties are very important and related with all fundamental physical properties, such as magnetism, superconductivity, thermodynamic, elastic, chemical bonding, hardness, and thermoelectric, properties. The half-metallic nature of materials is essential for spintronic device applications. The semiconducting materials are used for photovoltaic solar cell and thethermoelectric device. These nature of a material can be known by studying theelectronic structure of it. The calculated PBE-GGA energy bands of CoZrMnAl along with its total density of states are shown in the Fig. S2. Note that The valence bands largely cross the Fermi level for minority carriers but very small for majority carriers.

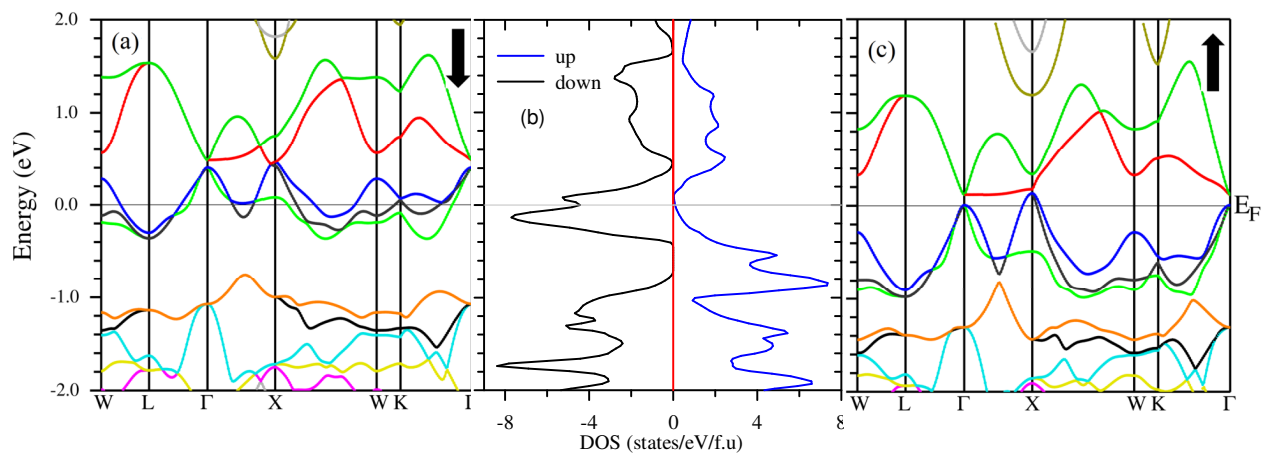

Fig.S2: Band structure and total density of states of CoZrMnAl including spin effect.

The total density of states for majority carriers at the Fermi level is very small. This is consistent with the calculated band structure. Thus, the material can be regarded as the "nearly half-metallic".

The calculated projected density of states of CoZrMnAl is shown in the Fig. S3. The high total density of states at the Fermi level for minority carriers arises from the Mn-3d states as shown in the Fig. S3 (e).

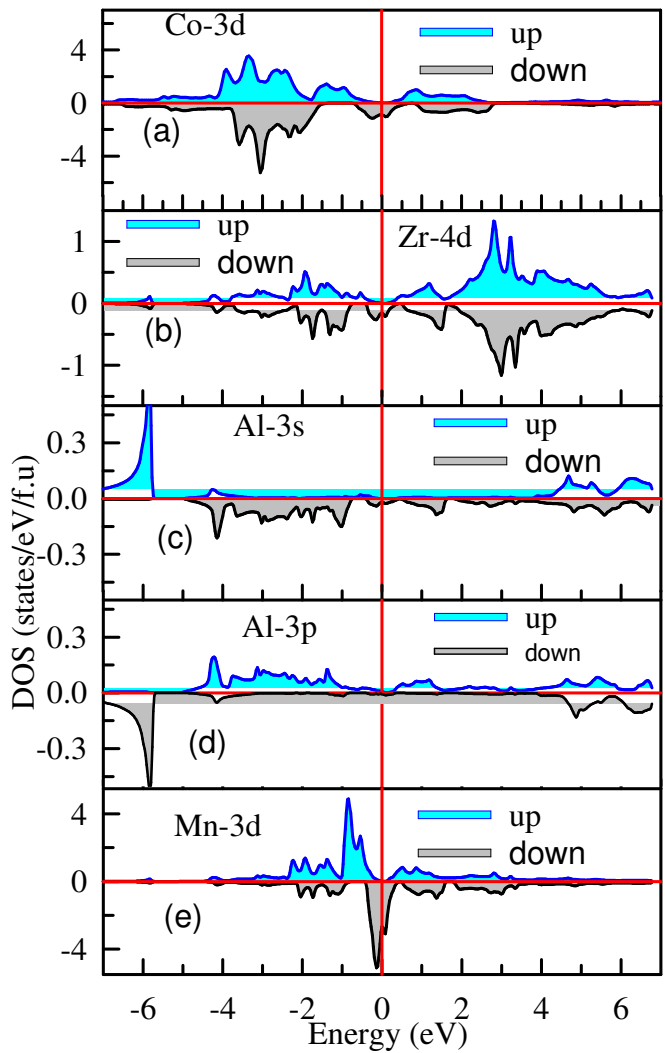

Fig. S3: Projected density of states of CoZrMnAl for both majority and minority carriers.

The energy band structure of CoZrMnGa and the total density of states for both majority and minority carriers are shown in the Fig. S4. The band structure shows metallic behavior for both spins. However, for majority carriers, the overlapping is small and thus the density of states at the Fermi level is smaller than that for minority carriers. Below the Fermi level, a large pseudogap exists for minority carriers. This arises from the Co-3d states which can be seen from the calculated projected density of states as shown in the Fig. S5.

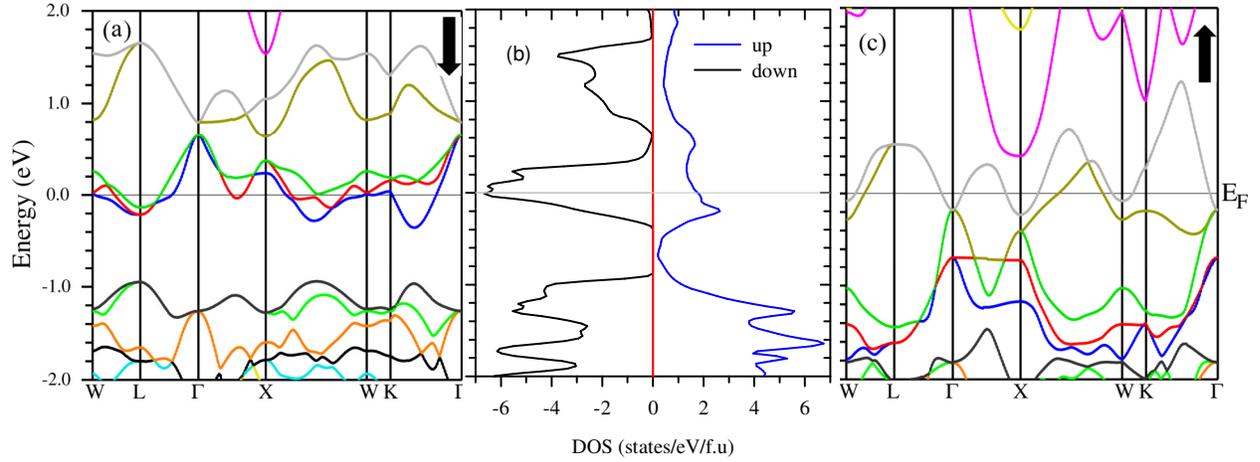

Fig. S4: Spin dependent band structure and total density of states of CoZrMnGa.

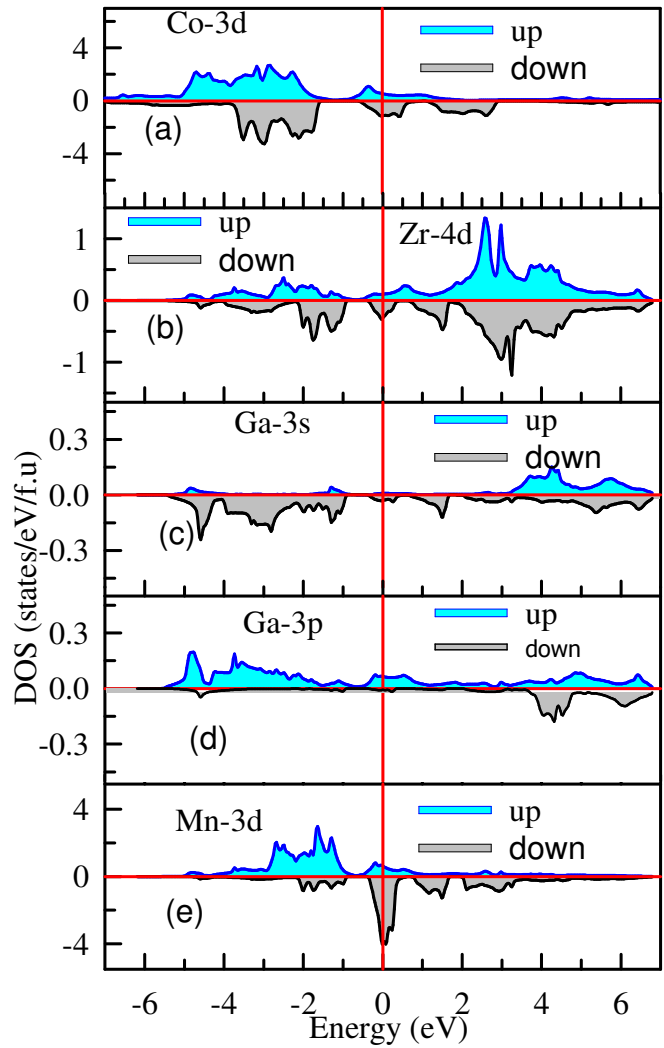

Fig.S5: Spin dependent projected density of states of CoZrMnGa.

The calculated energy bands and total density of states of CoZrMnGe are illustrated in the Fig. S6. The band structure and density of states for both majority and minority carriers are identical. A small bandgap 0.042 eV between Γ and X-points exists for both carriers and thus the material can be regarded as a narrow indirect bandgap semiconductor.

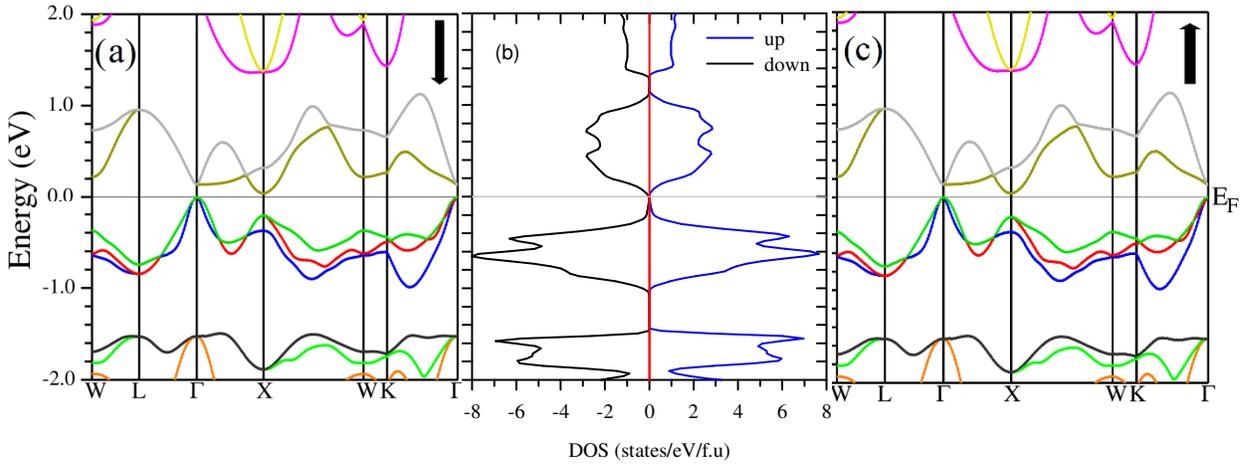

Fig.S6: The energy band structure and total density of states of CoZrMnGe including spin-effect.

The semiconducting nature of CoZrMnGe may arise due to the small contributions to the total density of states of Ge-4s for majority carriers, Ge-4p for minority carriers. This can be seen from the calculated projected density of states of CoZrMnGe as shown in the Fig. S5. Both majority and minority carriers of Co-3d, Zr-4d, and Mn-3d states equally contribute to the density of states, i. e., the projected density of states for both majority and minority carriers is equal. However, For Ge-4s and Ge-4p orbitals, this is not true as shown in the Fig. S5 (c) and (d). The band structure and total density of states for spin-up and spin-down are illustrated in the Fig. S6. A large number of valence bands cross the Fermi level for minority carriers but two balance bands and one conduction band cross the Fermi level for majority carriers. Thus, the total density of states at the Fermi level for minority carriers is comparatively smaller than that for minority carriers. A large pseudogap exists below the Fermi level for minority carriers. This arises from the Co-3d states as shown in the Fig. S9. Thus, the material shows metallic nature.

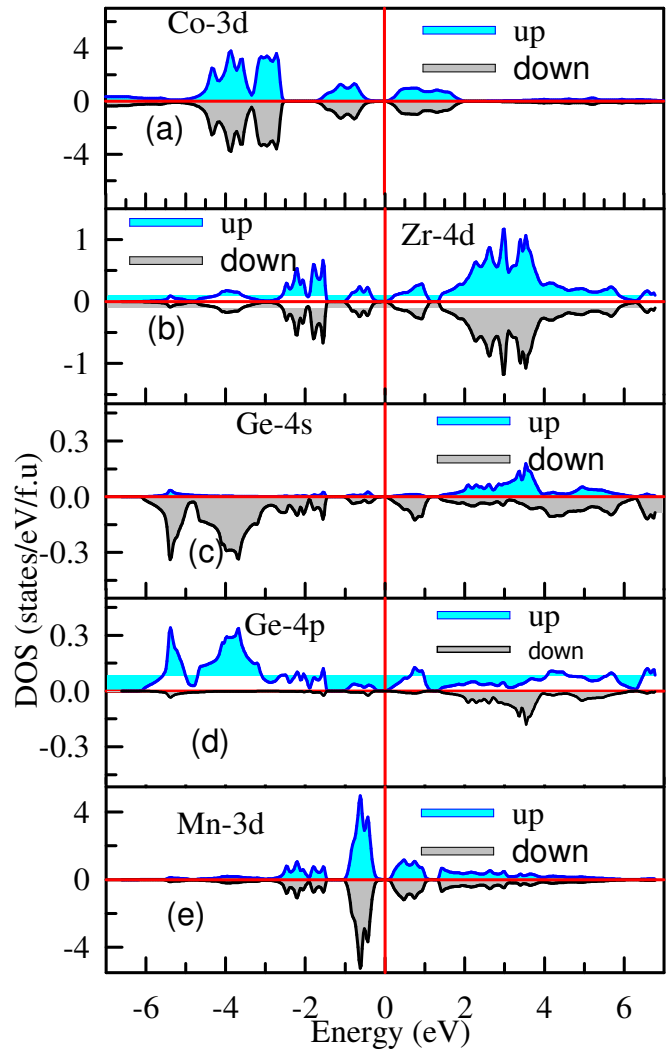

Fig.S7: Projected density of states of CoZrMnGe for both majority and minority carriers.

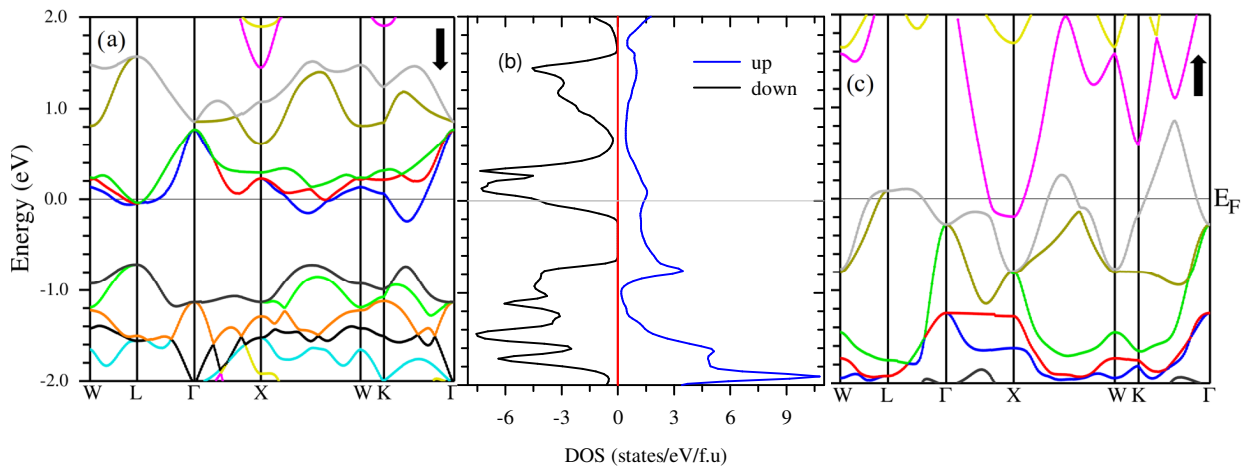

Fig. S8: The calculated spin-dependent band structure and total density of states of CoZrMnIn.

The calculated projected density of states (PDOS) of CoZrMnIn is shown in the Fig. S9. The most of the contributions to the total density of states come from the Co-3d, Mn-3d and Zr-4d states for both majority and minority carriers. The Mn-3d states for spin-down the prominent contributions to the total density of states (DOS) at the Fermi level (see Fig. S9 (e). The Co-3d states for minority carriers have more contributions at the Fermi level than that for majority carriers as shown in the Fig. S9 (a). This is also true for Zr-4d states (see Fig. S9 (b). The In-3s states have very small contributions to the Fermi energy, mostly from majority carriers. However, In-4 states exhibit opposite nature, have also small contributions to the total density of states at the Fermi level but for minority carriers as illustrated in the Fig. S9 (d) and (e).

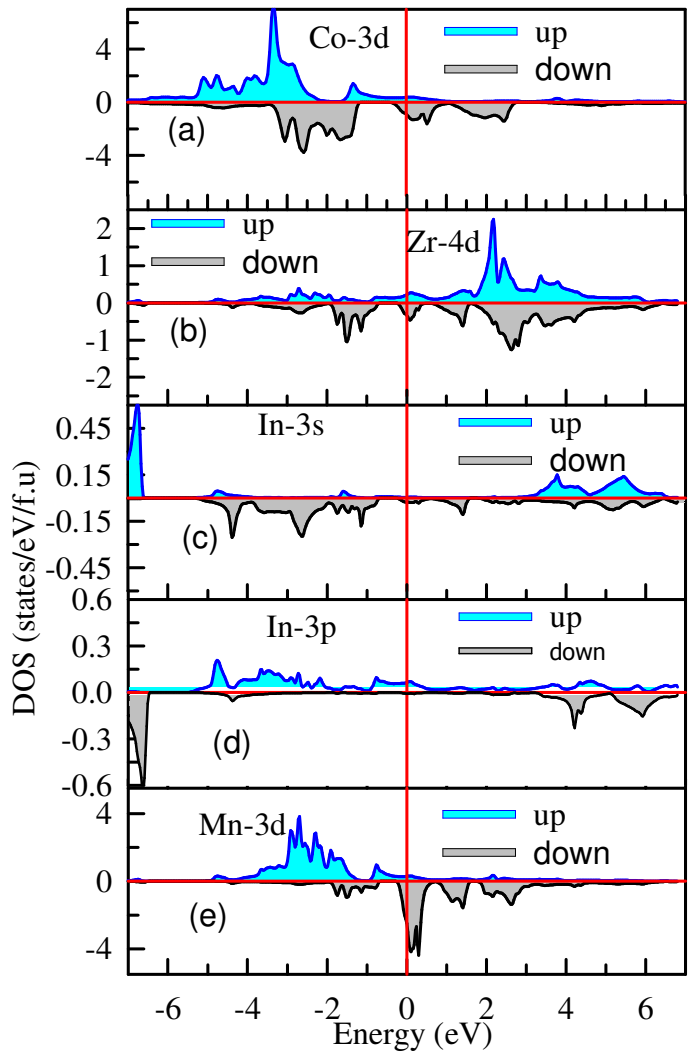

Fig.S9: Spin dependent projected density of states of CoZrMnIn.

For themain description of these properties, see the main article.